\documentstyle[12pt]{article}
\title{The Observational Appearance of Slim \\
Accretion Disks}
\author{Ewa Szuszkiewicz$^1$, Matthew A. Malkan$^{2}$, \\
 and Marek A. Abramowicz$^{3}$}
\date {}
\begin {document}
\baselineskip 12pt
\maketitle
\begin{center}
$^1$ International School for Advanced Studies, SISSA,\\
via Beirut 2-4, I-34013 Trieste, Italy \\
ewa@sissa.it
\end{center}
\begin{center}
$^2$ Department of Astronomy and Astrophysics, University of California, \\
Los Angeles, CA 90095-1562, USA\\
malkan@bonnie.astro.ucla.edu
\end{center}
\begin{center}
$^3$Department of Astronomy and Astrophysics, G{\"o}teborg
University \\
and Chalmers University of Technology, 412 96 G{\"o}teborg, Sweden \\
marek@fy.chalmers.se
\end{center}
\vspace{3.0cm}
\begin{center}
SISSA Ref. 109/95/A (September 1995)
\end{center}
\vspace{2.0cm}
\begin{center}
Accepted by Astrophysical Journal
\end{center}
\begin{center}
\newpage
{\bf Abstract}
\end{center}
We reexamine the hypothesis that the optical/UV/soft X-ray
continuum of Active Galactic
Nuclei 
is thermal emission from an accretion disk.
Previous studies have shown that fitting  the spectra with
the standard, optically thick and geometrically thin accretion
disk models 
often led to luminosities which
contradict the basic assumptions adopted in the standard model.
There is no known reason why the accretion rates in AGN should not be
larger than the thin disk limit. In fact, more general, slim accretion
disk models are self-consistent even for moderately super-Eddington
luminosities.
We calculate here spectra from a set of thin and slim, optically thick
accretion disks, assuming for simplicity a modified black body local emission
with no relativistic corrections.
We discuss the differences between the
thin and slim disk models, stressing the implications of these differences
for the interpretation of the observed properties of AGN.
We found that the spectra can
be fitted not only by models with a high mass and a low accretion rate
(as in the case of thin disk fitting) but also by models with a low mass
and a high accretion rate. In the first case fitting the observed spectra
in various redshift categories gives black hole masses around $10^9$
$M_{\odot}$ for a wide range of redshifts,
and for accretion rates ranging from 0.4 (low redshift) to 8 $M_{\odot}$/year
 (high redshift). In the second case the accretion rate is around $10^2$
$M_{\odot}$/year
for all AGN and the mass ranges from 3$\cdot 10^6$ (low redshift)
to $10^8$ $M_{\odot}$ (high redshift).
Unlike the disks with a low accretion rate, the  spectra
of the high-accretion-rate disks extend into the soft X-rays.
A comparison with  observations shows that such disks could produce
the soft X-ray excesses claimed in some AGNs.
We show also that the sequence of our models with fixed mass and
different accretion rates can explain the time evolution of the observed
spectra in Fairall 9.
\vspace{0.5cm}
\noindent

{\it Subject headings: active galactic nuclei - spectra - accretion
disks}

\section {Introduction}

The conjecture that Active Galactic Nuclei (AGN) are powered by
accreting supermassive black holes
(Lynden-Bell 1969) has not yet received a convincing
proof. Theoretical arguments based on the efficiency
of energy release, rapid
variability and superluminal expansion (Blandford 1992) strongly
support this hypothesis. However, the mechanism by which the black hole
powers the AGN is still unknown. There are at least three
classes of models for this mechanism: spherical accretion,
disk accretion and dynamo models (Price 1991).
In particular physical situations they can be fairly distinct, as for
example in disks powered by small scale turbulent viscosity, which may
not be driven by a dynamo. In other cases they are closely related.

Although ground- and space-based data have improved considerably,
even the most advanced observational techniques still are not able to resolve
the innermost parts of AGN. Therefore, accretion flow close to the
central engine (within 0.1 pc from the center)  can be studied only
indirectly.
Additional complications are introduced by possible reprocessing of the
central radiation out at larger radii.

In this paper we discuss a particular aspect of the accretion paradigm for
AGN---namely the predictions of slim accretion disk models for the
observational appearance of an accreting nonrotating black hole.
We use here  non self-gravitating optically thick disk
 models.
In Section 2  properties
of optically thick accretion disk models and
the results of previous studies of AGN spectra are summarized.
In Section 3 the comparison between structure and spectra of thin
and slim disk models  is
presented. General properties of slim disk spectra together with
the observed AGN spectra are subject of Section 4. Variability is
discussed in Section 5 and we conclude our paper
with a summary in Section 6.

\section{Spectra of AGN in the framework of accretion disks}

One of the most energetically dominant
observed features in the continuum spectrum of  quasars
is the ``Big Blue Bump" (hereafter BBB), extending from the optical to
the ultraviolet, rising
above extrapolations of the infrared continuum. Variability studies
suggest that this may be
a separate component from the infrared since it varies much more strongly.
The beginning of the bump is marked by an inflection around  a well-defined
wavelength of 1 $\mu$m. 
How far it extends into EUV or soft X-ray is not known yet.
This BBB has  most often been interpreted as thermal emission
from an optically thick geometrically thin accretion disk.
Following a promising suggestion by Shields (1978),
Malkan \& Sargent (1982) and Malkan (1983) showed that the strong
BBB of most QSOs and luminous
Seyfert 1 galaxies can be well-fitted with
simple models of geometrically thin Keplerian disks.
Already at that time, two objections to thin disk models were
discussed. They seem to predict too high linear polarization of
the UV continuum (e.g., Webb \& Malkan 1986)
or too strong a Lyman limit jump (Sun \& Malkan 1987).
The geometrically thin accretion disk model for QSO has also been criticized
on the ground that it requires luminosities near or in excess of
the Eddington luminosity so that it is not self-consistent.
Recently, synchronous optical-UV variability has been observed for a
few Seyfert galaxies (NGC 5548 is the best example). This may
contradict a thin disk prediction
that  most of the optical thermal continuum is emitted from larger
radii (where the temperature is lower)
than  most of the UV continuum.
The observational arguments mentioned above
are discussed by e.g. Malkan (1992),  Kinney (1994), Collin - Souffrin (1994).
Sophisticated accretion disk models have been calculated and
alternatives have been proposed, which have major weaknesses of their own.
Although the issue is not free from controversy, it is fair
to conclude that, after considerable observational and
theoretical effort, optically thick accretion models remain
a leading possible explanation for the Big Blue Bump.

In many sources, the Einstein IPC and EXOSAT satellites observed
ultrasoft (0.3 keV and below) excess flux above extrapolations of the
hard X-ray power law (Arnaud et al. 1985, Wilkes \& Elvis 1987, Turner
\& Pounds 1989, Masnou et al. 1992, Comastri et al. 1992).
The soft X-ray excess is generally better fitted by the curved, thermal
spectra models
rather than a second steep power law component (Urry et al. 1989, Weaver 1993).
There are suggestions that it is the
same physical component as the ultraviolet BBB. Some more recent
ROSAT PSPC spectra do not seem to confirm this, as they are mostly well
fitted by a single power-law
(Laor, Fiore, Elvis, Wilkes \& McDowell 1994).
A more reliable interpretation of the shape of the soft X-ray
continuum requires observations with high signal-to-noise ratios,
good energy resolution, and coverage of a broad energy range, which
has been possible, for example, in some simultaneous ROSAT/GINGA
observations.
Unfortunately the connecting extreme-UV wavelengths are blocked
from direct observation by
the large Galactic
opacity.
One alternative is to observe the ultraviolet spectra of very
high-redshift quasars (Reimers et al. 1992), however this  is
complicated by the large number of intervening high-z absorption
systems with significant optical depth at the Lyman limit.
The other alternative is to go the other side of
the Galactic opacity barrier and to observe low redshift quasars
in very soft X-rays (e.g., Laor et al. 1994).
In either case, the detailed shape of EUV/soft X-ray
excess---essential for testing the models---is known only roughly.

\subsection{Properties of optically thick accretion disks}

Theoretically computed  spectra of  the thermal radiation
emitted by accretion disks depend mainly on two physical
parameters: accretion rate of mass supplied to
the disk from outside,
$\dot M$, and the mass of the central black hole, $M$.
Observations determine the absolute luminosity and shape
of the energy distribution. To compare them
with the calculations we introduce a few useful definitions.
The total luminosity, $L$, of the accreting disk,
is related to $\dot M$ through:
\begin{equation}
L=\eta \dot M c^2,
\end{equation}
where $\eta $ is the efficiency at which the gravitational energy is
converted into radiation.
It has been rigorously shown that the efficiency of a stationary
disk  accreting onto a black hole equals the binding energy of
fluid elements at the inner
edge of a disk, $r_{in}$. For the pseudo-Newtonian potential,
which mimics the case of Schwarzschild black hole (Paczy\'nski
\& Wiita 1980),
assuming that the matter falls from infinity,
\begin{equation}
\eta = {1 \over 4}{r-2 \over (r-1)^2}\ , \  \ \
 r={r_{in} \over r_G},
\end{equation}
where $r_G=2GM/c^2$ is the Schwarzschild radius,  $c$ is the speed of
light and $G$ is the gravitational constant.

It is convenient to express $L$ and $\dot M$ in critical units.
The most common practice is to take for a critical luminosity the Eddington
limit, which is defined as the maximal
 luminosity of a spherical star in hydrostatic
equilibrium:
\begin{equation}
L_E = {4\pi c G M \over \kappa} = {2.5 \cdot 10^{38} \over 1+X}
{M \over M_{\odot}}
\rm {erg/ s},
\end{equation}
where X is the hydrogen content by mass and it is assumed that the
opacity $\kappa $ is due to electron scattering only.
For the critical accretion rate,  there are two equally
often used definitions. The first, which is used in this paper,
is called simply the critical accretion rate and has the form:
\begin{equation}
\dot M_C =
{L_E \over (\eta)_{max}c^2}.
\end{equation}
The advantage of this formula is that for sub-Eddington accretion
rates $\dot M/\dot M_{C}=L/L_E$. It is less convenient for super-Eddington
accretion rates.
For the pseudo-Newtonian potential $(\eta)_{max}$ is equal to
1/16, as the maximum of binding energy is attained at the
marginally stable orbit, $r_{ms}=3r_G$. In this case
\begin{equation}
\dot M_C={64\pi G M \over \kappa c} =
{4.45\cdot10^{18} \over 1+X}{M\over M_{\odot}}
\rm g/s.
\end{equation}
Assuming
$\eta_{max} =1$ (100\% efficiency), yields
the definition of the Eddington accretion rate:
\begin{equation}
\dot M _E={L_E \over c^2}=
 {2.78\cdot 10^{17}\over 1+X}{ M\over M_{\odot}} {\rm g/s}.
\end{equation}
For the Eddington accretion rate  $\dot M/\dot M_E=\eta^{-1}(L/L_E)$.

The $L-\dot M$ relation for slim accretion disks is
shown in Figure 1. It was obtained
by constructing a sequence of
self-consistent models (see Section 3) for a wide
range of $\dot m =
\dot M/\dot M_C$ and  fixed mass $m=M/M_{\odot}=10^8$.
The  disk luminosities were calculated from
the local flux $F$,
\begin{equation}
 L=2\int_{r_{in}}^{r_{out}}F 2\pi rdr .
\end{equation}
Figure 1 shows that for $\dot m \ge 1$
the luminosity rises only logarithmically
with accretion rate.
This was first found by Jaroszy\'nski, Abramowicz \& Paczy\'nski (1980),
Paczy\'nski
\& Wiita (1980) and Paczy\'nski (1980)
in the context of thick accretion disks.
Abramowicz, Czerny, Lasota \& Szuszkiewicz (1988) showed
that it is also the property of
slim accretion disks around stellar mass black holes. The $L-\dot M$
relation depends very weakly on the mass of the central object, so in fact
the solid curve in Figure 1 is similar to that in Abramowicz et al. (1988),
with the difference that it now extends
to higher accretion rates. However, this relation depends significantly
on the form and value of the viscosity.
The  dependence on both mass and viscosity are connected with the
finite size of the disk, constrained by the requirement that the self-gravity
of the disk is not important.
The solid line in Figure 1 was obtained
assuming that the  viscous stress tensor,
$\tau _{r\varphi}$, is
proportional to the total pressure and the dotted
line is for $\tau_{r\varphi}$ proportional to the effective pressure
defined by $\sqrt{PP_{g}}$ (see equation (8)).
In both cases the proportionality
coefficient, $\alpha $,  is taken equal to 0.001.

The luminosity---accretion rate relation has  important
implications for the interpretation of the observed properties of AGN.
Depending on their location in the $\dot M$, $M$ plane, accretion disks
have quite different physical properties.
Recently, Chen, Abramowicz, Lasota, Narayan \& Yi (1995) have demonstrated
that
all possible thermal equilibria
of accretion disks with an arbitrary cooling mechanism (including advection)
and with an arbitrary optical depth  are four physically different classes.
The optically
thick disks with small viscosity
discussed in this paper are Type I of their four classes of models.
This class  uses the concepts of
thin, slim
and thick disks:

a) {\it Thin accretion disks ($\dot m \le 0.2$)}.
The relative vertical thickness is very small, $H/r \ll 1$, rotation
is described by the Kepler law, $\Omega =\Omega _K$. The pressure gradient
in the horizontal direction and accretion velocity are dynamically
unimportant. There is a local heat balance: heat generated by viscous stress
is radiated away at the same radius through the disk surface.
These models are described quite accurately by the analytic Shakura-Sunyaev
solution. (For  references see Frank, King \& Raine 1992).

b) {\it Slim accretion disks (wide range of $\dot m$)}.
The relative thickness $H/r \le 1$ need not to be small. Rotation differs
slightly but importantly from the Keplerian one. The pressure gradient in the
horizontal direction is dynamically important.
Slim disk models are described by a
set of ordinary differential equations and
one must explicitly solve
the eigenvalue problem connected with the regularity condition at the sonic
radius (which does not coincide with the radius of the marginally stable
Keplerian orbit). Heat transport in the horizontal direction (by advection)
is an important cooling mechanism.
Only numerical solutions are
known. (For references see Abramowicz, Czerny, Lasota \& Szuszkiewicz 1988).

c) {\it Thick accretion disks ($\dot m \gg 1$) }.
The relative disk thickness is large, $H/r \sim 1$, and the disk shape
is toroidal, with narrow long funnels along the rotation axis. Rotation
is highly non-Keplerian.  Because of mathematical
complications, the internal physics properties are difficult to study
and are known only qualitatively. Surface properties, relevant for the
continuum spectra, are described by analytic (but phenomenological)
formulae. (For references see Abramowicz, Calvani \& Madau 1987).

For small accretion rates ($\dot m$ not greater than
approximately 0.2)  the structure
of a disk can be described
by thin disk
model  everywhere  except in the innermost region
(Section 3).
For higher accretion rates
several assumptions necessary in the construction of the
thin disk model are not valid.
For  $\dot m $ exceeding 1,   disks become radiation
pressure-supported with electron scattering as the dominant source of opacity.
They share many characteristics of the geometrically thick accretion
disks even if they do not have toroidal structure.
The division between slim and thick disks occurs when the
approximation of vertical integration breaks down. A
detailed comparison between one- and two-dimensional models is needed to derive
a proper range for slim disk model applicability (Papaloizou
\& Szuszkiewicz 1994).

\subsection{Fitting observed spectra}

Most previous investigations have compared observations of the BBB
with simple thin disk models (type a above).
However, several problems with that approach,
mentioned at the beginning of Section 2, motivated us in this paper
to  compare
the observations with slim disk models (type b).

The first motivation is that the slim disk
models  cover a wide range of accretion rate regimes which may be
more relevant to most quasars.
Within this range there is no sub-Eddington limit, as for thin disk models.
We illustrate this in Figure 2 which shows a large number of values
of $\dot M$ and $M$ obtained
from previous fitting to AGN spectra. The investigators,
who  performed these fittings, assumed that the
observed BBB is thermal emission from a standard {\it thin}
accretion disk.
For those quasar spectra
which were fitted by Kerr disk models, we applied the appropriate
efficiency  and mass corrections so that all the numbers in Figure 2
refer to Schwarzschild black holes, using the transformation relations
found by Sun \& Malkan (1989). They showed
that in going from an extreme Kerr to a Schwarzschild black hole,
$\dot M$ decreases by 2.5 times, while $M$ increases by
2.5 times. 
The empty squares illustrate the results  obtained by Laor (1990).
We have chosen only those objects from his list which were fitted also
by Sun \& Malkan
(1989), to avoid the comparison of two different data samples,
although there is no
indication that the results would  differ significantly for other
samples. The filled
squares are the Sun \& Malkan (1989) results with the inclination angle,
$i$, taken as in Laor (1990). The results of the fitting performed
by Malkan (1983), Wandel \& Petrosian (1988),
 Czerny \& Elvis (1987), Sun \& Malkan (1988), Band \& Malkan
(1989), Wandel (1991) and Tripp, Bechtold, Green (1994) are also presented.
Most of the objects are located
in the wide band of accretion rates, between  $\dot M$ equal to 0.01
$M_{\odot}$/year
and $\dot M$ equal to 100 $M_{\odot}$/year, governed by the slim disk models.
This particular sample includes only a few objects  with high
redshift. Most of them have $z<2.0$.
It is evident from Figure 2 that many  of these fits found values of
the parameters $\dot M$, $M$ which are
outside the range of the applicability of the thin disk theory.
Note that all (with one exception) Laor's (1990) points lay well within
thin disks range. This is because in his calculations the requirement
to satisfy the thin disk approximation was built-in.
For quasars with higher redshifts (and very high luminosities)
Wandel \& Petrosian (1988) derived super-Eddington luminosities.
The same was found by Tripp et al. (1994) for a substantial
portion of their sample. Padovani
(1989), using a completely different approach, came to the same conclusion.
Adding the high-redshift quasars to the sample
makes our argument in  favor of the slim disk models even stronger.

A second motivation comes from the observations which suggest that
the BBB may be  part of a very broad feature that also extends into the
soft X-rays (Arnaud et al. 1985; Czerny \& Elvis 1987).
 The dashed arrows in Figure 2 show the changes in
the values of $\dot M $ and $M$ after including in
the fitting procedure the soft
X-ray continuum (Laor 1990).

Finally, as discussed in Section 5,
the stability and variability properties of thin disks may
make them poorer models for most AGN spectra than slim
disks.
According to the thin disk theory models with
$\dot m >0.01$ and the standard $\alpha$-viscosity are thermally
unstable in their
innermost parts. Some non-standard viscosity prescriptions are consistent
with models which are everywhere stable.

\section{The structure and  spectra of thin and slim disks}

\subsection{The structure of disk models}
Slim disk models take advantage of the simplification due to vertical
integration as in the case of the standard thin disk models, but at the same
time they correctly describe important physical effects which dominate
the flow for $\dot m>0.2$, but which are omitted in the standard models.
The momentum equation for slim disks retains the inertial term,
$v_r(dv_r/dr)$, describing the dynamical importance of the accretion velocity,
$v_r$, and the horizontal pressure gradient, $\rho^{-1} (dP/dr)$.
The advective, horizontal heat flux, $v_rT(dS/dr)$ is included in the
energy equation. The remaining equations are the same as the Shakura-Sunyaev
ones.
The pseudo-Newtonian potential (Paczy\'nski \& Wiita 1980) is used to
describe the gravitational field of the central black hole. The inner boundary
condition uses the fact that there is no viscous torque across the horizon
of the black hole, while the outer boundary condition states that at
large radii the model of the flow is similar to that of Shakura-Sunyaev
models.
We consider disks of Population I composition
X=0.7, Y=0.27, Z=0.03. The opacities are taken from the tables of Cox
\& Stewart (1970).
We construct
a set of self-consistent models characterized by the mass of the central
object $m=M/M_{\odot}$, rate of accretion $\dot m$ and viscosity
parameter $\alpha $ to compare with observed  AGN properties.
The models use a
  family of   different
viscosity prescriptions
 in which the relevant
viscous stress tensor component is proportional to the following
combination of the total, $P$, and gas, $P_g$, pressures:
\begin {equation}
\tau_{r\varphi}=-\alpha P^{1-\mu /2}P_g^{\mu /2},
\end {equation}
where $\mu$ is a measure of the relative importance of $P$ and $P_g$ in
the process of viscous energy generation, and it ranges from
0 ($\tau_{r\varphi}$ proportional to the total
pressure) to 2 ($\tau_{r\varphi}$ proportional to the gas pressure).
We discuss here two interesting cases when $\mu =0$ and 1.

We require that the mass of the disk is small in comparison
with the mass of the central black hole, so the self-gravity is not
important. This sets an upper limit on the
accretion rate and, because of the accretion rate---luminosity relation,
also an upper limit on the luminosity for  given values of $m$ and
$\alpha $. The outer radius is determined
by the condition that local self-gravitational instabilities are not
present. These instabilities may develop when the disk density exceeds
$M/r^3$ where $M$ and $r$ are the mass of the central object and
the radius of the disk respectively. The problem
of thermal stability is treated separately in the Section 5.
We use only those models which, checked a posteriori, are
effectively optically
thick. This means that the geometric mean of the total, $\kappa$, and
free-free,
$\kappa_{ff}$, opacities multiplied by the surface
density, $\Sigma$, satisfy the
condition
\begin{equation}
\tau _{eff} ={1 \over 2}[\kappa_{ff}\kappa]^{1/2}
\Sigma >1
\end{equation}

The slim disk models discussed in this paper are determined by the
above assumptions and
by differential equations which describe mass, energy and momentum
conservation (see Abramowicz et al. 1988).
 The  structure equations for  a simple geometrically thin,
optically thick accretion  disk
can be obtained from the slim disk equations by making appropriate
approximations (Szuszkiewicz 1990).

Figure 3 shows the radial structure of the sequence of slim models with
different luminosities. We have chosen a disk with
black hole mass $10^8M_{\odot}$, $\alpha$ = 0.001 and $\mu$ = 0.
Two solid lines illustrate the changes in the location of the
inner and outer edges of the disks. It can be seen that for $\dot m<1$
the inner edge is very close to $r_{ms}=3r_G$. At $\dot m=1$ it changes
abruptly  and then tends to $r_{mb}=2r_G$. The radial extent of
the disk, defined as the radius where its self-gravity  is
comparable to
the gravity from the black hole,
reaches a minimum for $\dot m$ around 0.1.

A region of radiation-pressure dominance is situated above the
dashed line, and the region of
electron-scattering dominance is above the dotted line.
In the case of $\tau_{r\varphi} \propto P$ ($\mu=0$) the models become
optically thin in the very inner part of the disk for $1<\dot m<50$
(hatched region).
For higher accretion rates the surface density
increases and the disk is again optically thick.
The model with accretion rate $\dot m=0.001$ is gas-pressure
dominated ($P_{r} \ll P_{g}$). The main opacity source is free-free
absorption. It is  a ``clean" example of the  Shakura
\& Sunyaev (1973) ``outer" region of the disk.
Matter under the conditions described above will radiate as a
blackbody. In the  model with $\dot m =0.01$, there are several different
regions:
the innermost ($r<3.3r_G$) where $P_g>P_{r}$, $\kappa_{ff} > \kappa_{es}$,
then the narrow region ($3.3r_G<r<4r_G$) where $P_g>P_{r}$, $\kappa_{es}
> \kappa_{ff}$, next ($4r_G <r<43r_G$) where $P_r>P_g$, $\kappa_{es}>
\kappa_{ff}$, then ($43r_G<r<60r_G$) where $P_g>P_r$,
$\kappa_{es} > \kappa_{ff}$, and finally the
outermost $r>60r_G$ where $P_g>P_r$
and $\kappa_{ff} > \kappa_{es}$ as in the innermost one.
For higher accretion rates the innermost region shrinks towards the
central object and the outermost shifts further away. The whole
disk  becomes radiation-pressure dominated with  electron scattering
as the principal opacity source. This is the clean ``inner" Shakura-Sunyaev
region.
For bigger black hole masses  the disk reaches the
sequence of states just described for lower luminosities.

In the case of $\mu=1$, Figure 3  looks quite similar. The important
differences
are that
the size of the disk of a given luminosity
becomes smaller and  that
the optically thin region of the disk  is not
present for any value of $\dot m$.

The agreement between the structure and radiative flux of the disk
calculated by the slim model and
the thin one is satisfactory for the range of accretion rates where
the thin approximation holds.
This fact justifies usage of thin disk approximation for small
accretion rates to evaluate the properties of the flow, such as the density,
temperature or local flux, apart from their innermost regions.
However,
thin accretion disks with higher luminosities were used
to interpret the AGN observations, and  it is
interesting to see how they differ from the slim models.  We call thin
disk models with parameters outside the range of applicability
``extrapolated thin disks".
In Figure 4 we show the local flux, temperature,
and surface density for  two models from the sequence just described
with $\mu =0$ and $\dot m =0.5$ and 1. For higher accretion rates
up to $\dot m=50$, models have a very small optically thin part. Following
our self-consistency argument we will not consider those models further.
Instead,
in Figure 5 we present the properties of slim and extrapolated thin accretion
disk models for $\mu= 1$. Both the structure and local flux
 show significant differences, especially for  super-Eddington
luminosities.
For the critical accretion rate we
calculated the structure and flux for the extrapolated thin disk using
both asymptotic expressions for the region characterized by $P_r\gg P_g$
and $\kappa_{es}\gg\kappa_{ff}$ and numerical solutions of the
thin disk equations.
Taking into
account the very narrow gas-pressure dominated region prevents
singular behavior of the surface density.

Super-Eddington models are purely radiation-supported structures (for
example the model with $\dot m =5$ in Figure 5)
where the only opacity source is electron scattering and the
flux is determined by the local effective gravity.
In this case the heat flux from the surface may be easily estimated,
by analogy to supermassive spherical stars, as
\begin{equation}
F={c \over \kappa}g
\end{equation}
where $g$ is the effective surface gravity.
Our  super-Eddington models have their fluxes equal to the critical one.
This is the feature they share
with the particular type of thick accretion disk model
constructed by Paczy\'nski
(1980). Because of our self-consistency criteria, our models cannot
give arbitrarily large luminosities, and cannot extend arbitrarily far
away from the central black hole. In Figures  6 and 7
the maximum luminosity---mass and radius---mass
relations for two different definitions
of $\tau _{r\varphi}$
are shown respectively.
The disks with $\mu =1$ are smaller in size and less luminous than those with
$\mu =0$.
Comparing these results with similar ones obtained for the
toroidal shaped thick disk (Abramowicz, Calvani \& Nobili 1980)
we can conclude that slim disks cannot
reach such high luminosities as thick disks (100 $L_E$) but they can easily
produce moderately super-Eddington luminosities (10 $L_E$).
This can be understood in terms of significant differences in the rotation
law (angular momentum distribution).

\subsection{The radiation spectrum from the accretion disk}

Accretion rate and efficiency determine total
luminosity.
The spectrum of the radiation emitted from the disk surface
depends on its structure
and surface temperature, $T_S$.
This temperature plays a similar role as effective temperature in stars.
To calculate the spectra we need to know how and where
the energy is released.
The simplest assumption to make about the emergent spectrum is that
it is emitted locally at the rate prescribed by viscous energy transport.
The local spectrum of the thermal radiation may be one
of three typical distributions: a  blackbody, a modified blackbody
(where the coherent scattering is important) or a Wien spectrum (where
inverse Compton may be important).
For an effectively optically thick disk, we assume that the local spectrum
is a blackbody
or modified blackbody (Rybicki \& Lightman 1979), given by
\begin{equation}
I_{\nu} = {2B_{\nu }(T) \over  1+(1+\kappa_{es}/\kappa_{ff}(\nu))
^{1/2}}\label{inu}
\end{equation}
where $B_{\nu}$ is the Planck spectrum and
$$ \kappa_{es}=0.20(1+X)$$
$$\kappa_{ff}(\nu) \simeq 0.75\cdot 10^{25} \rho T^{-3.5}x^{-3}(1-e^{-x})
(1+X)(X+Y)$$
where $x=h\nu/kT$ and Gaunt factor is taken equal to 1.
For each disk annulus,
we calculated the modified blackbody spectrum, which takes into account the
dominance of electron scattering over absorption
solving equation (\ref{inu}) together with equation for $T_S$
\begin{equation}
F(r)=\pi \int_{0}^{\infty}I_{\nu}(T_S,\rho)d\nu
\end{equation}
where spectral intensity $I_{\nu}$ depends on the opacity, the density
and the temperature of the disk and $F(r)$ is the local energy release
at a particular radius $r$.

The overall spectrum is computed by integrating the local
spectra over the radial
extent of the whole disk
$$F_{\nu} =2\pi \int_{r_{in}}^{r_{out}} I_{\nu}rdr $$

Figure 8  shows the spectra radiated from the thin disk with $\dot m=0.1$
and the extrapolated thin disk with $\dot m=1$ for two different forms
of viscosity $\mu =0$ (solid line) and $\mu =1$ (dotted line).
The main difference in the optical is due to the fact that the more dense,
$\mu=1$, model has a smaller size, as it becomes self-gravitating closer
to the central object.
The Figure shows the
classic thin disk result ($\dot m =0.1$) where the middle of the
spectrum approximates
an $L_{\nu} \propto \nu ^{+1/3}$ power law.
At higher frequencies (in the ultraviolet), this power law
is steepened to a slope of roughly zero, by electron scattering,
which modifies the emitted local blackbody spectrum.
The differences between
modified blackbody and blackbody spectra are  especially significant
for high accretion rates, as it is shown in Figure 9.
The comparison of the actual slim  and extrapolated
thin disk spectra is
made in Figure 10.
Within the valid range  of the thin disk approximation, i.e.
for $L<0.2L_{E}$, the slim disk gives the same spectrum as the
thin one, as expected. Moreover the small
differences in local flux seen in Figures 4 and 5
for sub-Eddington models do not appear in their total
spectra at all.  Only for super-Eddington models is the difference
significant (see Figure 10b). It is particularly pronounced in
the EUV, so it is very important to be able to observe quasars at these
wavelengths.
In the optical or UV the difference between the two spectra are so small
that it is impossible to determined  systematically  how the
disk parameters $m$ and $\dot m$ would change if we applied the slim disk
models instead of the thin ones.

\section{Spectra of AGN }

\subsection{Spectra of AGN in three different redshift bands}

 We  wish to compare slim disk models with the broad range of observed
energy distributions of Seyfert 1 nuclei and quasars.
Our starting point is the large database of optical/ultraviolet spectra for
80 AGNs observed  with IUE and ground-based telescopes described by
Malkan (1988), and used in Zheng \& Malkan (1993).
Since we are not attempting to produce optimized fits to each individual
spectrum, we have adopted two statistical procedures to isolate the
spectrum of the BBB from these energy distributions, by subtracting the
continuum component which dominates at infrared wavelengths.
In the first alternative, following Sun \& Malkan (1989) we suppose that the
infrared continuum is described by a power law of slope -1.3 which
continues into optical wavelengths. The normalization for all objects
was determined by assuming that one quarter of the observed continuum
at 5300 {\AA} in every quasar arises
from the power law component.
This was the average value found in the large quasar sample fitted by
Sun and Malkan, which is a subset of the data we show here.
Alternately, we suppose that the infrared continuum is thermal emission
from dust grains with temperatures ranging up to the sublimation
temperature.  We describe their combined emission spectrum
with the analytic formulation developed by Malkan (1992):
$L_\nu \propto \nu^{-0.7} \times \exp (-\nu / \nu_0)$
with $\nu_0 = 10^{14.06}$ Hz.
Again we adopted a single normalization of the infrared component
for every quasar, assuming it produces 90 percent of the observed flux
at $\nu_0$, the characteristic average value in fits presented in
those papers.  As discussed there, the dust continuum leads to a
significantly higher inferred BBB flux in the red, as can be seen in
Figures 11, 12 and 13.

In these Figures we show the energy distributions of the AGN
in their rest frame divided into
three subsets according to redshift:
low with $z<0.15$ (Figure
11), intermediate with $0.15<z<0.7$ (Figure 12), and high
with $z>0.7$ (Figure 13).
The first panel of
each Figure is a collection of the  data obtained from
the observed spectra by subtracting the power law, the second shows
the same data after subtracting thermal dust emission and
the third one
gives the result of comparison between them. The observational points
which belong to the same object are not connected, because we are not
concerned at this point with fitting individual spectra. The line
drawn through the cloud of points is the average of the
AGN luminosities for a given frequency.
The substantial contribution of Balmer  continuum
and blended Fe II lines is present in the spectral region around
log $\nu$ =14.85-15.1 making the so called ``little bump".
The difference between BBB shapes that would
have been found had we made individual fits to each quasar spectrum
is smaller than this systematic difference that depends on
our assumed shape of the infrared component.

Next we compare the mean luminosities and the shapes of the BBB described
above with the expectations deduced from the slim
accretion disk models. We start from models with the parameters
suggested by the previous studies where thin disk models were used.
That is,  we choose black hole masses in the
range $10^8-10^{10} M_{\odot}$ and
accretion rates below the critical one.
For simplicity we will not discuss inclination effects and all spectra
presented in this Section were calculated with cos $i$=0.5.
Generally the models cannot account for the extremely flat energy
distribution in the case where a thermal dust infrared
component was removed, particularly
those with $\mu =1$, as they are steeper in the optical range.
For the infrared power-law subtraction, reasonable fits are obtained.
The objects
with low redshifts have a very flat energy distribution. The model with
$\mu =0$,
$m=8 \cdot 10^8$ and $\dot m =0.01$ gives the right
luminosity but
has some difficulties reproducing the shape of the spectrum (Figure 14).
Similar fits can be obtained by $\mu=1$, $m=10^9$ and $\dot m =0.01$.
It should be stressed however that in this
particular case it is difficult to determine the observed
energy distribution due to additional effects like starlight which must be
taken into account.
The intermediate redshift objects
are fitted quite well with the model  $\mu=0$, $m=8 \cdot 10^8$ and
$\dot m =0.08$ (Figure 15) or with $\mu=1$, $m=10^9$ and
$\dot m =0.07$. Similarly the observations
of high  $z$ quasars favor
the model with $\mu=0$, $m=10^9$ and $\dot m =0.2$
or $\mu =1$, $m=2\cdot 10^9$ and $\dot m =0.15$ for the
power law-subtracted  spectra.
The best fit for $\mu =0$ together with
an attempt to fit dust-corrected spectrum with the model $\mu =0$,
$m=1.5\cdot 10^9$ and $\dot m =0.2$
is shown in  Figure 16.

Flat energy distributions suggest the possibility that slim
disk models with smaller masses and super-Eddington luminosities can
provide better fits to the observations. We explore this suggestion and
the results are presented in Figures 14, 17 and 18. It is very difficult
to find a satisfactory fit for low luminosity objects. Their optical flux
appears to come from the outer parts of the disk which are
self-gravitating. The very high accretion
rates necessary for obtaining such flat spectra could suggest high starburst
activity, as there is enough material for stars to be born.
Another more likely explanation is that the majority of the optical
flux is {\it reprocessed} EUV radiation intercepted on its way out
from the inner disk.
For intermediate objects
very high accretion rates are also required. The best fit 
is for high redshift objects.
It is interesting to notice that higher redshift (and thus higher luminosities)
require smaller dimensionless accretion rates---the opposite of claims
from fitting the previous set of thin disk models.

\subsection{Soft X-Ray excess}

The  low mass - high accretion rate models discussed above
extend further into the soft X-rays and can  produce
the observed excess component in this waveband, while
the high mass-low accretion rate models can be responsible only for the optical
and UV part of the spectrum. Here we examine the hypothesis that
the soft X-ray excess observed in an AGN originates in an
optically thick accretion disk.

We have selected
three particularly well observed bright, low-redshift
AGN (Figure 19) with a high soft X-ray flux relative to their optical
emission.
Other objects have less pronounced soft-X-ray excesses, while
still others show no excess at all.
However, the strong soft X-ray-excess AGN are exactly the objects
which present the greatest challenge to accretion models.
Our choice was largely based on EINSTEIN
data.
The observations of the bright Seyfert 1 nuclei in
Markarian 509 and 841, and the extreme soft-X-ray excess quasar PG 1211+14,
are taken from Band \& Malkan (1989) - crosses - in Figure 19.
And we have added the most recent observations by ROSAT - power law with
a bowtie - from Walter \& Fink (1993) and joint ROSAT/GINGA data -
open squares - from Pounds et al (1994). In general these new data,
which provide better energy resolution,
give a less dramatic soft X-ray excess.
The X-ray spectra were not, in general, measured simultaneously with
the longer wavelength data.  Nonetheless, possible X-ray variability
is not so large that it has obscured the evidence that the spectral
turnup in the softest X-rays could come from the high-frequency tail
of the BBB.

In Figure 19a we show the fit to the spectrum of Markarian 509 with a
disk model which extends up to 0.2 keV (solid
line). The disk parameters in this case have the following values:
$\dot m = 0.94$ and $m=5\cdot 10^7$.
For comparison we  also present  a fit to
the BBB with $\dot m=50$ and $m=8\cdot 10^6$  which extends
up to 1 keV (dotted line).
In the case of the Markarian 841, the least luminous of the
three AGN considered here (Figure 19b), the corresponding values are $\dot m=
0.5$, $m=2.5\cdot 10^7$ (solid line) and $\dot m = 10^3$, $m=10^6$ (dotted
line).
For PG 1211+143 the masses and the accretion rates derived from the fitted
models are $\dot m =1.5$, $m=5\cdot 10^7$ (solid line) and $\dot m=
100$, $m=10^7$ (dotted line).
Extreme soft X-ray fluxes
require according to our models very high accretion rates,
in absolute units,
particularly in the case of the broad disk component in Mkn 841,
which may be unlikely. Relativistic correction and rotation of
the black hole, which we plan to implement in future studies may give
harder spectra at lower accretion rates.

The model spectra fall steeply (exponentially) in the soft X-ray regime,
while the majority of observed spectra are less steep (see e.g. Laor et al.
1994). This may indicate that the flux from the optically
thick accretion disk extends no further than
0.5 keV. A new generation of accretion disk models with the appropriate
treatment of the optically thin part of the flow will provide a more
definitive conclusion.

\section{Variability}

As was shown in Section 3, the thin disk model,
within the limits of its validity, gives a reasonable approximation to the
observed properties of accretion onto a compact object. The stability
problem requires taking into account  all details of the accretion
process which in the case of the black hole differ significantly
from those of accretion onto other compact objects. This can be considered only
in the slim approximation since the thin disk assumption discards information
about the inner part of the flow which is most relevant for stability.
Stability depends very strongly on the viscosity assumption.
One way of extracting information
about both properties of AGN and the viscosity mechanism itself is to
compare predictions from the models with different forms of viscosity with
the observed properties of AGN.
Here we apply the  family of
viscosity prescriptions given in Equation (8).

Abramowicz, Czerny, Lasota \& Szuszkiewicz
(1988) suggested a limit cycle instability can explain
the variability of X-ray sources.  If this mechanism is
present in AGN, one can expect to see not only intensity variations,
but also spectral variability,
i.e., a varying BBB.
A plausible upper limit to the amplitude of variability comes from
comparing the flux from the lower stable state of the disk with the flux
from the upper stable state.  The standard viscosity prescription gives
disks with such wide-ranging instability that extremely large flux
variations (around three orders of magnitude) would appear to be allowed.
With the modified viscosity law, the amplitudes are reduced to one or
one and a half orders of magnitude, more consistent with the
largest variations seen in normal AGN.
This is shown in Figure 20 where the
stability regions (from a simple local analysis) of the thin and slim
disks in the $\dot m - r$ plane
are drawn.
Similar studies were performed for slim disks around
stellar black holes (Honma, Matsumoto \& Kato 1991; Szuszkiewicz 1990).
The global analysis by Honma et al. (1991) and local analysis by
Wallinder (1991) suggests that the disk is
stabilized before reaching the upper stable branch of the slim
disk solution so that the unstable inner part is smaller in the $\dot m$
direction. These results still await confirmation.

We illustrate how this mechanism could work in Figure 21.
The difference
in the UV flux (for example at 1544 {\AA})
between low and high states is about a factor 20, as was observed in
Fairall 9 (Clavel, Wamsteker, \& Glass 1989).
We have plotted the three characteristic energy distributions for
this high-luminosity Seyfert 1 nucleus, corresponding to maximum,
minimum and average brightness levels.
The IUE data are from Clavel et al. (1989)
with simultaneous optical photometry from Glass (1986).
As with the quasars, we have corrected the optical fluxes by
subtracting an infrared component attributed to either a power law,
or thermal dust emission.  Even after this correction, it is apparent
that the range of variation in the UV (a factor of ~10) is
greater than the range in the red (a factor of 3.5).
This illustrates the typical trend for AGN spectra to become harder
when brighter.

Let us first discuss the spectrum corrected for thermal dust emission.
We will use here models with $\mu=1$, as they fit data better.
The smaller BBB
at the luminosity minimum is taken as the energy distribution
arising from the disk
with $m=2\cdot 10^8$ and
$\dot m =3.8\cdot 10^{-2}$.  At maximum light, the spectrum
is represented by the model with the same mass and the
$\dot m =0.65$. The intermediate state is fitted by
the model with $\dot m=0.09$.
In the case of the spectrum corrected for the power law the best fits
are obtained using spectra calculated in the simple sum-of-blackbodies
approximation. The sequence of models fitting the observations has
mass $m=2\cdot 10^8$ and $\dot m =0.03$, 0.08, and 0.6 for
minimum, intermediate and high states.
The instability operates in the inner part of the disk (see Figure 20).
Predicted thermal instability timescales for
the variability is of the order of 1000$^d$, depending on $\alpha$.
This picture is in rough
agreement with the proposed mechanism for variability.

It is intriguing that slim disks tend to stabilize at the highest accretion
rates.  This might suggest that the most luminous AGN should show
little UV variability.
At intermediate accretion rates, for sub-Eddington disks,
we might expect the BBB and the soft-X-ray
excess to vary more or less directly together.
In highly super-Eddington disks, the UV spectrum could remain steady while the
X-ray spectrum brightened and hardened.  (Note that in these optically
thick models, no Comptonization takes place).
However, our investigations presented here are too approximate
to make strong conclusions about detailed spectral variability properties.
This will require calculation of the full time evolution for
such models (Miller \& Szuszkiewicz 1995).

Another interesting feature of supercritical models
is that they give similar  radiated energy distributions
from disks with different accretion rates. The reason
is that for higher accretion rates the heat trapped in matter becomes
important, and the flow of matter induces non-negligible advective
horizontal heat flux. Thus, for higher accretion rates, efficiency
decreases and luminosity increases not in proportion to the accretion
rate but more slowly (Figure 1).

\section{Conclusions and discussion}

Within the range  of the validity of the thin disk approximation, i.e.
for $L<0.2L_{E}$, the slim disk reproduces all characteristics of the
thin disk, as we expected. Moreover the slim disk extends
the standard model to higher luminosities, which might be
necessary to explain the
luminous quasars.
The slim disk models provide the stabilizing mechanism over a
range of accretion rates.
The super-Eddington models
 offer the same quality of fits to the observations as the sub-Eddington ones
and have the advantage of also being able to explain the soft X-rays spectra.

It is difficult to compare our results with previous studies, mostly
because they differ in essential points. We start from the self-consistent
slim disk models and treat the spectra in a very simplified way. The
others start from very simplified disk model and include important
physical effects in the calculations of the spectra. Moreover they
used a much bigger viscosity parameter, which at higher accretion rates
requires proper treatment of the optically thin part of the disk.

In the case of a geometrically thin accretion disk it is possible to
consider the vertical  and  radial disk structures separately. It is
also reasonable to assume local heat balance. In order to calculate
a spectrum it is sufficient to divide the disk into a number of
one-dimensional layers and to solve the radiative transfer there.
Many sophisticated accretion disk models have been calculated since
the first simplest local sum-of-blackbodies approximation.
The slim accretion disk model takes into account non local flux
generation, but still radiative transfer is considered only in
the vertical direction.
Solving radiative transfer only in the vertical direction, justified
when the disk is very thin, leads to the so called "lags" of the optical
continuum behind (or ahead of) the UV continuum (depending on the propagation
direction of the perturbation). Such models predicts long lags as the
optical radiation is generated far away from the place where the UV is
radiated, and the wavelike disturbances travel more slowly than light.
Quasi-simultaneous optical and UV spectra of NGC 5548 obtained by
a combined IUE and ground-based  monitoring campaign
(Clavel et al.  1991 and Peterson et al. 1991)
 show that the average lag of the
optical continuum behind the UV continuum is no larger than 4 days.
In NGC 4151 a similar limit is no larger than 3 days. In Fairall 9,
there are indications that an optical-UV lag has been detected and
it is 140 days. A similar detection was found for 3C 273 and the
estimated lag is no larger than 40 days.

When the disk is geometrically and optically
thick
local energy balance cannot be assumed, since the heat generated by
dissipation can travel in any direction before reaching the surface.
The energy produced at some point inside the disk  will not just
diffuse vertically but may  emerge from any part of the surface.
In such disk we do not expect to observe long lags.

Two very important improvements remain to be made. One is to
include self-consistently optically thin part of the disk.
We would like to mention here that there are many new developments
in the area of optically thin advection-dominated accretion flows
(e.g., Abramowicz, Chen, Kato, Lasota, Regev 1995; Narayan \& Yi 1995;
Chen et al. 1995; Bj\"{o}rnsson, Abramowicz,
Chen, Lasota 1995).
Second,
is to solve the radiative transfer equation in geometrically  and optically
thick accretion
disks.

\newpage
\parindent 0pt
\parskip 11pt
{\bf References}


Abramowicz, M. A., Calvani, M., \& Madau, P., 1987, Comments
Astrophys. 12, 67

Abramowicz, M. A., Calvani, M., \& Nobili, L., 1980, ApJ,
242, 772

Abramowicz, M. A., Chen, X., Kato, S., Lasota, J-P., \& Regev, O., 1995,
ApJ, 438, L37

Abramowicz, M. A., Czerny, B., Lasota, J-P., \& Szuszkiewicz, E., 1988,
ApJ 332, 646

Arnaud, K. A., et al., 1985, MNRAS 217, 105

Band, D. L., \& Malkan, M. A., 1989, ApJ 345, 122

Bj\"{o}rnsson, G., Abramowicz, M. A., Chen, X., \& Lasota, J-P., 1995,
ApJ, submitted

Blandford, R. D., 1992 in Physics of Active Galactic Nuclei,
eds. W. J. Duschl, S. J. Wagner, (Heidelberg: Springer-Verlag)

Chen, X., Abramowicz, A. M., Lasota, J-P., Narayan, R., \& Yi, I., 1995,
ApJ, 443, L61

Clavel, J. C., Wamsteker, W., \& Glass, I., 1989, ApJ 337, 249

Clavel, J. C., et al., 1991, ApJ 366, 64

Collin-Souffrin, S., 1994, in  Theory of Accretion Disks - 2,
eds, W. J. Duschl, J. Frank, F. Meyer, E. Meyer-Hofmeister, W. M.
Tscharnuter, (Dordrecht: Kluwer Academic Publishers)

Comastri, A., et al., 1992, ApJ 384, 62

Cox, A. N, \& Steward, J. N., 1970, ApJ Suppl. 19, 243

Czerny, B., \& Elvis, M., 1987, ApJ 321, 305

Frank, J., King, A. R., \& Reine, D. J., 1992, Accretion power
in Astrophysics, (Cambridge: Cambridge University Press)

Glass, I., 1986, MNRAS 219, 5P

Honma, F., Matsumoto, R., \& Kato, S., 1991, PASJ 43, 147

Jaroszy\'nski, M., Abramowicz, M. A., \& Paczy\'nski, B., 1980,
Acta Astron. 30, 1

Kinney, A. L., 1994, in The First Stromlo Symposium: The Physics
of Active Galaxies, ASP Conference Series 54, ed., G. V. Bicknell, A.
Dopita, and P. J. Quinn, in press

Laor, A., 1990, MNRAS 246, 369

Laor, A., Fiore, F., Elvis, M., Wilkes, B. J., \& McDowell, J. C., 1994,
ApJ 435, 611

Lynden-Bell, D., 1969, Nature 223, 690

Malkan, M. A., \& Sargent, W. L. W., 1982, ApJ 254, 22

Malkan, M. A., 1983, ApJ 268, 582

Malkan, M. A., 1988, Adv. Space Res. 8, 249

Malkan, M. A., 1992,
 in Physics of Active Galactic Nuclei,
eds. W. J. Duschl, S. J. Wagner, (Heidelberg: Springer-Verlag)

Masnou, J-.P, Wilkes, B. J., Elvis, M., Arnaud, K. A., \& McDowell, J. C.,
1992, a\&A 253, 35

Miller, J. C., \& Szuszkiewicz, E., 1995, in preparation

Narayan, R., \& Yi, I., 1995, ApJ, 444, 231


Paczy\'nski, B., 1980, Acta Astron. 30, 347

Paczy\'nski, B., \& Wiita, P. J., 1980, A\&A 88, 23

Padovani, P., 1989, A\&A 209, 27

Papaloizou, J. C. B., \& Szuszkiewicz, E., 1994, MNRAS 268, 29

Pounds, K. A., et al., 1994, MNRAS 267, 193

Peterson, B. M., et al., 1991, ApJ 368, 119

Price, R. H., 1991, in Annals New York Academy of Sciences 631,
p. 235, eds. J. R. Buchler, S. L. Detweiler and J. R. Isper, (New York:
New York Academy of Sciences)

Reimers, D., Vogel, S., Hagen, H.-J, Engels, D., Groote, D.,
Wamsteker, W., Clavel, J., \&  Rosa, M. R.,  1992, Nature 360,
561

Rybicki, G. B., \& Lightman, A. P., 1979, in Radiative Processes
in Astrophysics, (New York: Wiley)

Shakura, N. I., \& Sunyaev, R. A., 1973, A\&A 24, 337

Shields, G. A., 1978, Nature 272, 706

Sun, W. -H., \&  Malkan, M. A., 1987, in Astrophysical Jets and
Their Engines, ed. W. Kundt, (Dordrecht: Reidel)

Sun, W. -H., \&  Malkan, M. A., 1988, in Supermassive Black Holes, ed.
M. Kafatos, (Cambridge: Cambridge University Press)

Sun, W. -H., \& Malkan, M. A., 1989, ApJ 346, 68

Szuszkiewicz, E., 1990, MNRAS 244, 377

Tripp, T. M., Bechtold, J., \& Green, R. F., 1994, preprint

Turner, T. J., \& Pounds, K. A., 1989, MNRAS 240, 833

Urry, C. M., Arnaud, K. A., Edelson, R. A., Kruper, J. S., \&
Mushotzky R. F., 1989, in Proc. 23$^{rd}$ ESLAB Symp. on Two Topics in X-ray
Astronomy, eds. J. Hunt and B. Battrick, ESA SP-296, Vol 2, 789

Wallinder, F. H., 1991, A\&A 249, 107

Walter, R. \& Fink, H. H., 1993, A\&A 274, 105

Wandel, A., \& Petrosian, V., 1988, ApJ 329, L11

Wandel, A., 1991, A\&A 241, 5

Weaver, K. A. , 1993, Ph. D. Thesis, University of Maryland

Webb, W. \& Malkan, M. A., 1986, in The Physics of Accretion Onto
Compact Objects, eds. Mason, K. O., Watson, M. G. \& White, N. E.,
(Berlin: Springer-Verlag) p.15

Wilkes, B. J., \& Elvis, M., 1987, ApJ 323, 243

Zheng, W., \& Malkan, M. A., 1993, ApJ 415, 517
\newpage

{\bf Figure Captions}

\parindent 0pt

Figure 1: The total luminosity - accretion rate relation for slim
disk models with the two different forms of viscosity: $\tau_{r\varphi}
=-\alpha P$ (solid line), $\tau_{r\varphi} = -\alpha \sqrt{PP_g}$
(dotted line). In both cases $\alpha $ =0.001.

Figure 2:
The values of $\dot M  $ and $M$ obtained by fitting
the observed spectra to those predicted by theory:

$\Box $ Laor (1990) \\
$\Box$ Sun \& Malkan (1989) \\
$\times$ Sun \& Malkan (1988), Band \& Malkan (1989), Wandel (1991),
Czerny \\
\hspace{1.cm} \& Elvis (1987) \\
$\diamondsuit $ Padovani (1989) \\
$\ast$ Tripp et al. (1994)  \\
solid line rectangular - quasars - Wandel \& Petrosian (1988)\\
dotted line rectangular - Seyferts -  Wandel \& Petrosian (1988)

Figure 3: The radial structure of the sequence of models with
different luminosities. The mass of the central black hole is 10$^8M_{\odot}$,
$\alpha =0.001$, and $\mu=0$. The size of the disk is the distance
between two solid lines. The dotted line marks the place where
$\kappa_{es}=\kappa_{ff}$ and the dashed one where $P_g=P_r$. In the small
hatched region in the left corner  the models
become effectively optically thin.

Figure 4: Local flux (a), temperature (b), surface density (c)  for the thin
(dotted lines)
and slim disk models (solid lines) with $m=10^8$,
$\alpha =10^{-3}$, $\mu =0$ and
$\dot m=0.5$. Similarly (d), (e) and (f) for the $\dot m =1$.

Figure 5: Local flux (a), temperature (b), surface density (c)  for the thin
(dotted lines)
and slim disk models (solid lines) with $m=10^8$,
$\alpha =10^{-3}$, $\mu =1$ and
$\dot m=0.5$. Similarly (d), (e) and (f) for the $\dot m =1$. Figures
(g), (h) and (i) are also for $\dot m =1$ but instead of using the
numerical solution for standard thin disk model we used the asymptotic
Shakura-Sunyaev formula. Figures  (j), (k) and (l) show the properties
of the models with $\dot m =5$.

Figure 6: The maximum possible luminosity of non-self-gravitating
slim accretion disk orbiting a central black hole as a function
of its mass
for two different viscosity prescriptions: $\mu=0$ (solid
line), $\mu =1$ (dotted line).

Figure 7: The outer radius for the slim disks with maximal luminosity
for two different viscosity prescriptions: $\mu=0$ (solid
line), $\mu =1$ (dotted line).

Figure 8: The spectral distributions for the models with $\dot m=0.1$
and $\dot m = 1$ for different viscosity prescriptions: $\mu=0$  (solid
lines), $\mu=1$  (dotted lines).

Figure 9: The comparison between the spectra of the $\mu=1$ disks
with $\dot m=0.1$ and 1 calculated as the superposition of modified
blackbodies (dotted lines) and simple black bodies (solid lines).

Figure 10: The slim disk spectra  (solid lines) versus the extrapolated
thin one (dotted lines)
a) for $\dot m$=
0.5, 1 b) for $\dot m=$ 5.

Figure 11: Energy distributions for 23 quasars with $z<0.15$.
a) after correction for an infrared power law, b) after correction for
thermal infrared emission from hot dust, c) the difference between the average
distributions in the case a) and b).

Figure 12: Similarly as in Figure 11 but for 45
quasars with $0.15<z<0.7$.

Figure 13: Similarly as in Figure 11 but for 37
quasars with $z>0.7$.

Figure 14: The attempt to fit the low-redshift objects with
two models: first with $m=8\cdot 10^8$, $\dot m=0.012$ and second
with $m=3\cdot 10^6$, $\dot m=10^3$

Figure 15: The best fit to the intermediate-redshift objects:
$m=8\cdot 10^8$, $\dot m=0.08$

Figure 16: The best fits to the high-redshift objects:
for the power law-corrected spectra: $m=1.5\cdot 10^9$, $\dot m=0.2$,
and for the dust-corrected spectra: $m=2\cdot 10^9$, $\dot m=0.2$

Figure 17: The best fit to the intermediate-redshift objects:
for the power law-corrected spectra: $m=2\cdot 10^7$, $\dot m=250$,
and for the dust-corrected spectra: $m=2\cdot 10^7$, $\dot m=500$

Figure 18: The best fits to the high-redshift objects:
for the power law-corrected spectra: $m=8\cdot 10^7$, $\dot m=70$,
and for the dust-corrected spectra: $m=8\cdot 10^7$, $\dot m=130$

Figure 19: The best fits to the three low-luminosity AGN with
spectra observed into the soft X-rays a) MKN 509
- $m=5\cdot 10^7$, $\dot m =0.94$ (solid line), $m=8\cdot 10^6$,
$\dot m =50$ (dotted line),
b) MKN 841 - $m=2.5\cdot 10^7$, $\dot m =0.5$ (solid line),
$m=10^6$, $\dot m =10^3$ (dotted line), the dashed line shows the
spectrum for $m=10^6$, $\dot m =10^3$ if the disk structure is
extrapolated beyond the self-gravity point, c) PG 1211+143 -
$m=5\cdot 10^7$, $\dot m =1.5$ (solid line), $m=10^7$,
$\dot m =100$ (dotted line). See text for more details.

Figure 20: The stability regions in $\dot m - r$ space for thin (a)
and slim (b) accretion disks with $m=10^{8}$, $\alpha =0.001$ and different
 values of $\mu$.

Figure 21: The spectral evolution
 in the source Fairall 9 and the best fits for a) spectrum corrected for
an infrared power law b) spectrum corrected for thermal dust emission.

\end{document}